\documentclass[12pt]{article}
\usepackage{setspace}
\usepackage{amsmath}
\usepackage{amsfonts,amsthm}
\usepackage{latexsym}
\usepackage{amscd}
\usepackage{amssymb}
\usepackage{graphicx}
\usepackage{float}
\usepackage[applemac]{inputenc}
\usepackage{dsfont}
\usepackage{multirow}
\usepackage{dcolumn}
\usepackage{rccol}
\usepackage{enumerate}
\usepackage{scalefnt}
\usepackage{url}
\usepackage[titletoc]{appendix}

\usepackage[pdftex,colorlinks=true]{hyperref}
\hypersetup{colorlinks=true, linkcolor=red, filecolor=blue, pagecolor=blue, urlcolor=green, citecolor=blue}

\usepackage{textcomp}
\usepackage{array}
\usepackage{supertabular}
\usepackage{hhline}

\makeatletter
\newcommand\arraybslash{\let\\\@arraycr}
\makeatother
\newcolumntype{+}{>{\global\let\currentrowstyle\relax}}
\newcolumntype{^}{>{\currentrowstyle}}

\newlength{\bracewidth}

\def\fudge{\mathchoice{}{}{\mkern.5mu}{\mkern.8mu}}
\def\bbc#1#2{{\rm \mkern#2mu\vbar\mkern-#2mu#1}}
\def\bbb#1{{\rm I\mkern-3.5mu #1}}
\def\bba#1#2{{\rm #1\mkern-#2mu\fudge #1}}
\def\bb#1{{\count4=`#1 \advance\count4by-64 \ifcase\count4\or\bba A{11.5}\or
   \bbb B\or\bbc C{5}\or\bbb D\or\bbb E\or\bbb F \or\bbc G{5}\or\bbb H\or
   \bbb I\or\bbc J{3}\or\bbb K\or\bbb L \or\bbb M\or\bbb N\or\bbc O{5} \or
   \bbb P\or\bbc Q{5}\or\bbb R\or\bbc S{4.2}\or\bba T{10.5}\or\bbc U{5}\or
   \bba V{12}\or\bba W{16.5}\or\bba X{11}\or\bba Y{11.7}\or\bba Z{7.5}\fi}}

\restylefloat{float}

\title{Assessing the Efficiency of \textit{Cordon Sanitaire} as a Control Strategy of Ebola}

\author{Baltazar Espinoza, Victor Moreno, Derdei Bichara and Carlos Castillo-Chavez}

\date{}

\begin{document}

\maketitle

\begin{abstract}
We formulate a two-patch mathematical model for Ebola Virus Disease dynamics in order to evaluate the effectiveness of \textit{cordons sanitaires}, mandatory movement restrictions between communities while exploring their role on disease dynamics and final epidemic size. Simulations show that severe restrictions in movement between high and low risk areas of closely linked communities may have a deleterious impact on the overall levels of infection in the total population.
\end{abstract}
 
\textbf{keywords:} Ebola virus disease, Asymptomatic, Final size relation, Residence times.

\section{Introduction}
Ebola virus disease (EVD) is caused by a genus of the family \textit{Filoviridae} called \textit{Ebolavirus}. The first recorded outbreak took place in Sudan in 1976 with the longest most severe outbreak taking place in West Africa during 2014-2015 \cite{WHOEbola2014}. Studies have estimated disease growth rates and explored the impact of interventions aimed at reducing the final epidemic size \cite{ChoCast04,Legrand:2007ai,lekone2006statistical,TowersPatersonCCC14SIAM}. Despite these efforts, research that improves and increases our understanding of EVD and the environments where it thrives is still needed \cite{PetersLeDuc99}.\\
This chapter is organized as follows: Section 2 reviews past modeling work; Section three introduces a single Patch model, its associated basic reproduction number $\mathcal{R}_0$, and the final size relationship; Section four introduces a two-Patch model that accounts for the time spent by residents of Patch $i$ on Patch $j$; Section 5 includes selected simulations that highlight the possible implications of policies that forcefully restrict movement (\textit{cordons sanitaires});and, Section 6 collects our thoughts on the relationship between movement, health disparities, and risk.

\section{Prior Modeling Work}
Chowell et \textit{al.} \cite{ChoCast04} estimated the basic reproduction numbers for the 1995 outbreak in the Democratic Republic of Congo and the 2000 outbreak in Uganda. Model analysis showed that control measures (education, contact tracing, quarantine) if implemented within a reasonable window in time could be effective. Legrand et \textit{al.} \cite{Legrand:2007ai} built on the work in \cite{ChoCast04} through the addition of hospitalized and dead (in funeral homes) classes within a study that focused on the relative importance of control measures and the timing of their implementation. Lekone and Finkenst\"{a}dt \cite{lekone2006statistical} made use of an stochastic framework in estimating the mean incubation period, mean infectious period, transmission rate and the basic reproduction number, using data from the 1995 outbreak. Their results turned out to be in close agreement with those in \cite{ChoCast04} but the estimates had larger confidence intervals.
 
The 2014 outbreak is the deadliest in the history of the virus and naturally, questions remain  \cite{chowell2015modelling,chowell2014western,House:2014fk,NishiuraGC2014,Pandey2014,TowersPatersonCCC14SIAM,TowersPattCCC2014}.  Chowell et {\it al}. in \cite{chowell2015modelling} recently introduced a mathematical model aimed at addressing the impact of early detection (via sophisticated technologies) of pre-symptomatic individuals on the transmission dynamics of the Ebola virus in West Africa. Patterson-Lomba et {\it al}. in \cite{TowersPattCCC2014} explored the potential negative effects that restrictive intervention measures may have had in Guinea, Sierra Leone, and Liberia. Their analysis made use of the available data on Ebola Virus Disease cases up to September 8, 2014. The focus on \cite{TowersPattCCC2014} was on the dynamics of the``effective reproduction number'' $R_\textsl{eff}$, a measure of the changing rate of epidemic growth, as the population of susceptible individuals gets depleted. $R_\textsl{eff}$ appeared to be increasing for Liberia and Guinea, in the initial stages of the outbreak in densely populated cities, that is, during the period of time when strict quarantine measures were imposed in several areas in West Africa. Their report concluded, in part, that the imposition of enforced quarantine measures in densely populated communities in West Africa, may have accelerated the spread of the disease. In \cite{chowell2014western}, the authors showed that the estimated growth rates of EVD cases were growing exponentially at the national level. They also observed that the growth rates exhibited polynomial growth at the district level over three or more generations of the disease. It has been suggested that behavioral changes or the successful implementation of control measures, or high levels of clustering, or all of them may nave been responsible for polynomial growth. A recent review of mathematical models of past and current EVD outbreaks can be found in \cite{chowell2014transmission} and references therein. Inspired by these results, we proceed to analyze the effectiveness of forcefully local restrictions in movement on the dynamics of EVD. We study the dynamics of EVD within scenarios that resemble EVD transmission dynamics within locally interconnected communities in West Africa.

\section{The model derivation}\label{mod}
\textit{Cordons Sanitaire} or ``sanitary barriers'' are designed to prevent the movement, in and out, of people and goods from particular areas. The effectiveness of the use of \textit{cordons sanitaire} have been controversial. This policy was last implemented nearly one hundred years ago \cite{byrne2008encyclopedia}. In desperate attempts to control disease, Ebola-stricken countries enforced public health officials decided to use this medieval control strategy, in the EVD hot-zone, that is, the region of confluence of Guinea, Liberia and Sierra Leone \cite{NYT2014ebola}.
In this chapter, a framework that allows, in the simplest possible setting, the possibility of assessing the potential impact of the use of a \textit{Cordon Sanitaire} during an EVD outbreak, is introduced and ``tested''.
The population of interest is subdivided into susceptible ($S$), latent ($E$), infectious ($I$), dead ($D$) and recovered ($R$). The total population (including the dead) is therefore $N=S+E+I+D+R$. The susceptible population is reduced by the process of infection, which occurs via effective ``contacts'' between an infectious ($I$) or a dead body ($D$) at the rate of $\beta(\frac{I}{N}+\varepsilon\frac{D}{N})$ and susceptible. EVD-induced dead bodies have the highest viral load, that is, more infectious than individuals in the infectious stage ($I$); and, so, it is assumed that $\varepsilon > 1$. The latent population increases at the rate $\beta S(\frac{I}{N}+\varepsilon\frac{D}{N})$. However since some latent individuals may recover without developing an infection \cite{AsympEbola2000,SymptomLess2000,BellanPulliam,ChoCast04,hawryluck2004sars,EbolaGabon2005}, it is assumed that exposed individuals develop symptoms at the rate $\kappa$ or recover at the rate $\alpha$. The population of infectious individuals increases at the rate $\kappa E$ and decreases at the rate $\gamma I$. Further, individuals leaving the infectious stage at rate $\gamma$, die at the rate $\gamma f_{dead}$ or recover at the rate $1-\gamma f_{dead}$. The $R$ class includes recovered or the removed individuals from the system (dead and buried). By definition the $R$-class increases, the arrival of previously infected, grows at the rate $(1-f_\textrm{dead})\gamma I$.

\begin{figure}[H]
\centering
\includegraphics[scale=0.7]{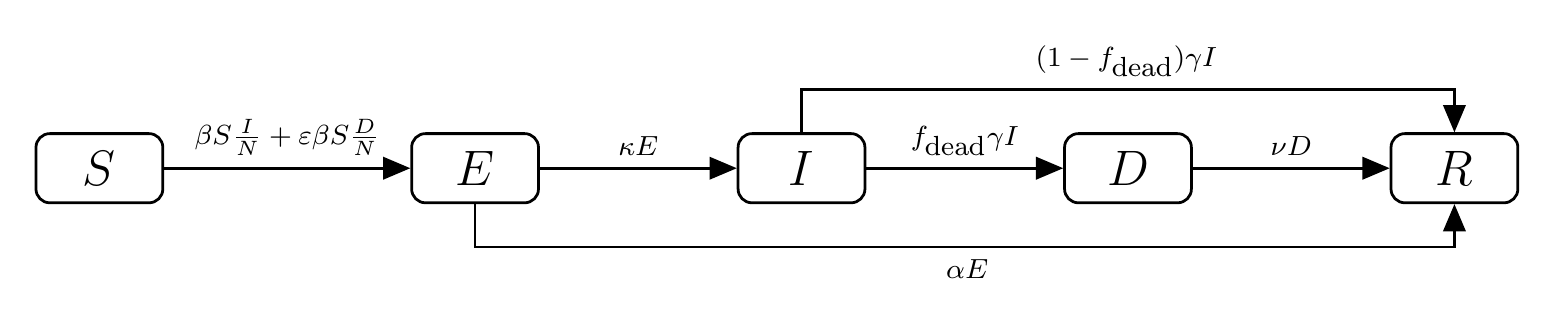}
\caption{An SEIDR Model for Ebola virus disease}
\label{Flow1}
\end{figure}

A flow diagram of the model is in Fig. \ref{Flow1}, The definitions of parameters are collected in Table \ref{tab:par}, including the parameter values used in simulations

where
\begin{table}[H]
\begin{center}
\begin{tabular}{llc}
\textbf{Parameter}&\hspace{3cm}\textbf{Description}& \hspace{0.3cm} \textbf{Base model values}\\
\hline
\hspace{0.7cm}$\alpha$			& Rate at which of latent recover without developping symptoms
& \hspace{0.5cm} $0-0.458$ \cite{AsympEbola2000} \\
\hspace{0.7cm}$\beta$			& Per susceptible infection rate & \hspace{0.5cm} $0.3056$ \cite{chowell2015modelling, TowersPattCCC2014, chowell2014transmission}\\
\hspace{0.7cm}$\gamma$		& Rate at which an infected recovers or dies & \hspace{0.5cm} $\frac{1}{6.5}$ \cite{chowell2014transmission}\\
\hspace{0.7cm}$\kappa$		& Per-capita progression rate & \hspace{0.5cm} $\frac{1}{7}$ \cite{chowell2015modelling, TowersPattCCC2014}\\
\hspace{0.7cm}$\nu$				& Per-capita body disposal rate & \hspace{0.5cm} $\frac{1}{2}$ \cite{Legrand:2007ai}\\
\hspace{0.7cm}$f_\textrm{dead}$	& Proportion of infected who die due to infection & \hspace{0.5cm} $0.708$ \cite{chowell2014transmission}\\
\hspace{0.7cm}$\varepsilon$	& Scale: Ebola infectiousness of dead bodies & \hspace{0.5cm} 1.5\\
\hline
\end{tabular}
\end{center}
\caption{Variables and parameters of the contagion model.}
\label{tab:par}
\end{table}
The mathematical model built from Fig. \ref{Flow1}, that models EVD dynamics is given by the following nonlinear systems of differential equations:

\begin{equation} \label{EbolaAsym}
\left\{\begin{array}{ll}
N=S+E+I+D+R\\\\
\dot S=-\beta S\frac{I}{N}-\varepsilon\beta S\frac{D}{N}\\\\
\dot E=\beta S\frac{I}{N}+\varepsilon\beta S\frac{D}{N}-(\kappa+\alpha) E\\\\
\dot I= \kappa E-\gamma I\\\\
\dot D=f_{\textrm{dead}}\gamma I-\nu D\\\\
\dot R=(1-f_{\textrm{dead}})\gamma I+\nu D+\alpha E
\end{array}\right.
\end{equation}
The total population is constant and the set $\Omega=\{(S,E,I,R)\in \mathbb{R}_{+}^4 / S+E+I+R\leq N \}$ is a compact positively invariant, that is, solutions behave as expected biologically. Hence Model (\ref{EbolaAsym}) is well-posed. Following the next generation operator approach \cite{MR1057044,VddWat02} (on $E$, $I$ and $D$), we find that the basic reproductive number is given by
$$
\mathcal R_0=\left(\frac{\beta}{\gamma}+\frac{\varepsilon f_{\textrm{dead}}\beta}{\nu}\right)\frac{\kappa}{\kappa+\alpha}
$$
That is, $\mathcal{R}_0$ is given by the sum of the secondary cases of infection produced by infected and dead individuals during their infection period. The final epidemic size relation that includes dead (to simplify the maths) being given by
$$\log\frac{N}{S^\infty}=\mathcal R_0\left(1-\frac{S^\infty}{N}\right).$$

\section{EDV dynamics in heterogeneous risk environments}\label{het}
\begin{figure}[H]
\centering
\includegraphics[scale=0.9]{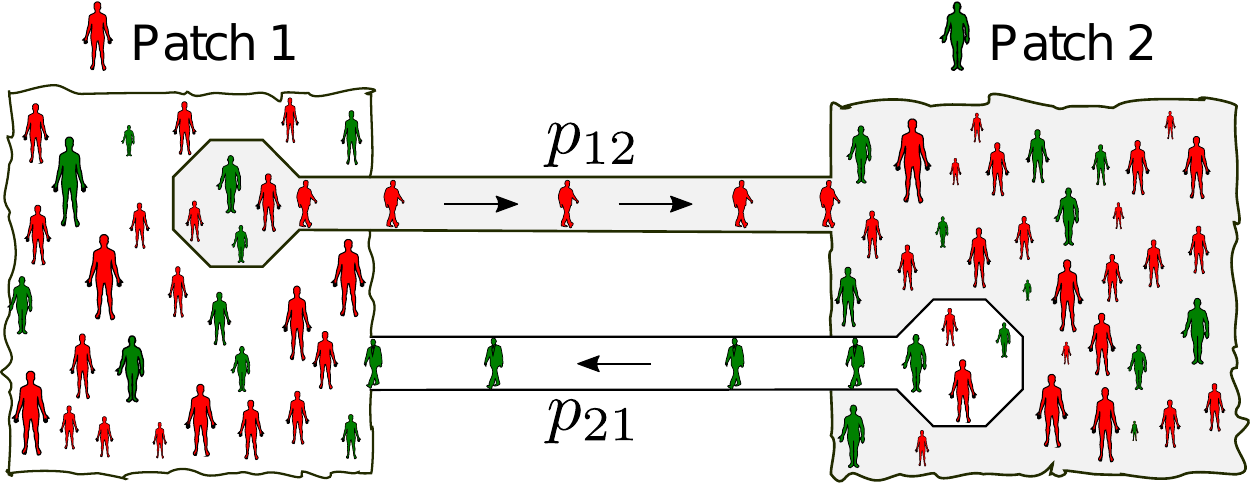}
\caption{Dispersal of individuals via a Lagrangian approach.}
\label{lagmov}
\end{figure}

The work of Eubank {\it et al.} \cite{eubank2004modelling}, Sara de Valle {\it et al.} \cite{stroud2007spatial}, Chowell {\it et al.} \cite{chowell2003scaling}, and \cite{bichara2015sis} analyze heterogeneous environments. Castillo-Chavez and Song \cite{castillo2003epidemic}, for example,  highlight the importance of epidemiological frameworks that follow a Lagrangian perspective, that is,  models that keep track of each individual (or at least its place of residence or group membership) at all times. The figure \ref{lagmov} represents a schematic representation of the Lagrangian dispersal between two patches.

Bichara {\it et al.} \cite{bichara2015sis} uses a general Susceptible-Infectious-Susceptible (SIS) model involving n-patches given by the following system of nonlinear equations:
 
$$
\left\{\begin{array}{ll}
\dot S_i=b_i-d_iS_i+\gamma_i I_i-\sum_{j=1}^n(\textrm{$S_i$ infected in Patch } j)\\
\dot I_i=\sum_{j=1}^n(\textrm{$S_i$ infected in Patch } j)- \gamma_i I_i-d_i I_i\\
\dot N_i=b_i-d_iN_i.
\end{array}\right.
$$

where $b_i$, $d_i$ and $\gamma_i$ denote the per-capita birth, natural death and recovery rates respectively. Infection is modeled as follows:
$$\left[ S_i\; \textrm{infected in Patch }j\right]= \underbrace{\beta_j}_{\textbf{the risk of infection in Patch $j$ }}\times \underbrace{p_{ij}S_i}_{\textbf{Susceptible from Patch $i$ who are currently in Patch $j$ }}$$ $$\times \underbrace{\frac{\sum_{k=1}^{n}p_{kj}I_k}{\sum_{k=1}^{n}p_{kj}N_k}}_{\textbf{Proportion of infected in Patch $j$}}.$$

where the last term accounts for the {\it effective} infection proportion in Patch $j$ at time. The model reduces to the single n-dimensional system

$$
\dot I_i=\sum_{j=1}^n\left(\beta_j p_{ij}\left(\frac{b_i}{d_i}-I_i\right) \frac{\sum_{k=1}^{n}p_{kj}I_k}{\sum_{k=1}^{n}p_{kj}\frac{b_k}{d_k}}\right)- (\gamma_i +d_i) I_i
\quad i=1,2,\dots,n.
$$

with a basic reproduction number $\mathcal R_0$  that it is a function of the risk vector $\mathcal B=\left(\beta_1,\beta_2,\dots,\beta_n\right)^t$ and the residence times matrix $\mathbb P=(p_{ij})$, $i,j=1,...,n$, where $p_{i,j}$ denotes the proportion of the time that an $i$-resident spends visiting patch $j$.
In \cite{bichara2015sis}, it is shown that when $\mathbb P$ is irreducible (patches are strongly connected),
 the Disease Free State is globally asymptotically stable if $\mathcal R_0\leq1$ (g.a.s.) while, whenever  $\mathcal R_0>1$ there exists a unique interior equilibrium which is g.a.s.

The Patch-specific basic reproduction number is given by
$$
 \mathcal R_0^i(\mathbb P)={\mathcal R}_0^i \times \sum_{j=1}^n\left(\frac{\beta_j}{\beta_i}\right) p_{ij}\left( \frac{\left(p_{ij} \frac{b_i}{d_i}\right)}{\sum_{k=1}^{n}p_{kj}\frac{b_k}{d_k}}\right).
 $$
  where $\mathcal R_0^i$ are the \textit{local} basic reproduction number when the patches are isolated. This Patch-specific basic reproduction number gives the dynamics of the disease at Patch level \cite{bichara2015sis}, that is, if  $\mathcal R_0^i(\mathbb P)>1$ the disease persists in Patch $i$.  Moreover, if $p_{kj}=0$ for all $k=1,2,\dots,n$ and $k\neq i$ whenever $p_{ij>1}$, it has been shown  \cite{bichara2015sis} that the disease dies out form Patch $i$ if $\mathcal R_0^i(\mathbb P)<1$. The authors in  \cite{bichara2015sis} also considered a multi-patch SIR single outbreak model and deduced the final epidemic size. The SIR single outbreak model considered in  \cite{bichara2015sis} is the following:
 $$
\left\{\begin{array}{ll}
\dot S_i=-\left(\frac{\beta_ip^2_{ii}}{p_{ii}N_i+p_{ji}N_j}+\frac{\beta_jp^2_{1ij}}{p_{ij}N_i+p_{jj}N_j}\right)S_iI_i    -  \left(\frac{\beta_ip_{ii}p_{ji}}{p_{ii}N_i+p_{ji}N_j}+\frac{\beta_jp_{ij}p_{jj}}{p_{ij}N_i+p_{jj}N_j}\right)S_iI_j,\\\\
\dot I_i=\left(\frac{\beta_ip^2_{ii}}{p_{ii}N_i+p_{ji}N_j}+\frac{\beta_jp^2_{1ij}}{p_{ij}N_i+p_{jj}N_j}\right)S_iI_i  + \left(\frac{\beta_ip_{ii}p_{ji}}{p_{ii}N_i+p_{ji}N_j}+\frac{\beta_jp_{ij}p_{jj}}{p_{ij}N_i+p_{jj}N_j}\right)S_iI_j            -\alpha_i I_i,\\\\
\dot R_i=\alpha_i I_i,
\end{array}\right.
$$

where $i,j=1,2, i\neq j$, and $S_i, I_i$ and $R_i$ denotes the population of susceptible, infected and recovered immune individuals in Patch $i$, respectively. The parameter $\alpha_i$ is the recovery rate in Patch $i$ and $N_i\equiv S_i+I_i+R_i$, for $i=1,2$.

In this chapter we will be making use of this modeling framework, but with a slightly different formulation, to test under what conditions the movement of individuals from high risk areas to nearby low risk areas due to the use of {\it cordon sanitaire}, is effective in reducing {\it overall} transmission by considering two-Patch single outbreak that captures the dynamics of Ebola in a two-patch setting.
\subsection{Formulation of the model}
It is assumed that the community of interest is composed of two adjacent geographic regions facing highly distinct levels of EVD infection. The levels of risk account for differences in population density, availability of medical services and isolation facilities, and the need to travel to a lower risk area to work. So, we let $N_1$ denote be the population in patch-one (high risk) and $N_2$ be the population in patch-two (low risk). The classes $S_i$, $E_i$, $I_i$, $R_i$ represent respectively, the susceptible, exposed, infectious and recovered sub-populations in Patch $i$ ($i=1,2$). The class $D_i$ represents the number of disease induced deaths in Patch $i$. The dispersal of individuals is captured via a Lagrangian approach defined in terms of residence times \cite{bichara2015sis,bichara2015VB}, a concept developed for communicable diseases for $n$ patch setting \cite{bichara2015sis} and applied to vector-borne diseases to an arbitrary number of host groups and vector patches in \cite{bichara2015VB}.   \\

We model the new cases of infection per unit of time as follows:
\begin{itemize}
\item The density of infected individuals mingling in Patch 1 at time t, who are only capable of infecting susceptible individuals currently in Patch 1 at time $t$, that is, the {\it effective} infectious proportion in Patch 1 is given by
$$p_{11}\frac{I_1(t)}{N_1} + p_{21}\frac{I_2(t)}{N_2},$$
where $p_{11}$ denotes the proportion of time residents from Patch 1 spend in Patch 1 and $p_{21}$ the proportion of time that residents from Patch 2 spend in Patch 1.
\item The number of new infections within members of Patch 1, in Patch 1 is therefore given by
$$\displaystyle \beta_1 p_{11}S_1\left(p_{11}\frac{I_1(t)}{N_1} + p_{21}\frac{I_2(t)}{N_2}\right).$$
\item The number of new cases of infection within members of Patch 1, in Patch 2 per unit of time is therefore
$$\displaystyle \beta_2 p_{12}S_1\left(p_{12}\frac{I_1(t)}{N_1} + p_{22}\frac{I_2(t)}{N_2}\right),$$
where $p_{12}$ denotes the proportion of time that residents from Patch 1 spend in Patch 2 and $p_{22}$ the proportion of time that residents from Patch 2 spend in Patch 2; given by the effective density of infected individuals in Patch 1 
$$p_{11}\frac{I_1(t)}{N_1} + p_{21}\frac{I_2(t)}{N_2},\quad (*)$$
while the \textit{effective} density of infected individuals in Patch 2 is given by 
$$p_{12}\frac{I_1(t)}{N_1} + p_{22}\frac{I_2(t)}{N_2}.\quad (**)$$
Further, since, $p_{11}+p_{12}=1$ and $p_{21}+p_{21}=1$ then we see that the sum of (*) and (**) gives the density of infected individuals in both patches, namely,
$$\frac{I_1}{N_1}+\frac{I_2}{N_2},$$
as expected. If we further assume that infection by dead bodies occurs only at the local level (bodies are not moved) then, by following the same rationale as in  Model \ref{EbolaAsym}, we arrive at the following model:
\end{itemize}
\begin{equation} \label{EboPatchy}
\left\{\begin{array}{ll}
N_1=S_1+E_1+I_1+D_1+R_1\\\\
N_2=S_2+E_2+I_2+D_2+R_2\\\\
\dot S_1=-\beta_1p_{11}S_1\left(p_{11}\frac{I_1}{N_1}+p_{21}\frac{I_2}{N_2}\right)-\beta_2p_{12}S_1\left(p_{12}\frac{I_1}{N_1}+p_{22}\frac{I_2}{N_2}\right)-\varepsilon_1 \beta_1p_{11}S_1\frac{D_1}{N_1}\\\\
\dot E_1=\beta_1p_{11}S_1\left(p_{11}\frac{I_1}{N_1}+p_{21}\frac{I_2}{N_2}\right)+\beta_2p_{12}S_1\left(p_{12}\frac{I_1}{N_1}+p_{22}\frac{I_2}{N_2}\right) +\varepsilon_1 \beta_1p_{11}S_1\frac{D_1}{N_1}-\kappa E_1-\alpha E_1\\\\
\dot I_1=\kappa E_1-\gamma  I_1\\\\
\dot D_1=f_{\textrm{dead}}\gamma I_1-\nu D_1\\\\
\dot R_1=(1-f_{\textrm{dead}})\gamma I_1+\nu D_1+\alpha E_1\\\\
\dot S_2=-\beta_1p_{21}S_2\left(p_{11}\frac{I_1}{N_1}+p_{21}\frac{I_2}{N_2}\right)-\beta_2p_{22}S_2\left(p_{12}\frac{I_1}{N_1}+p_{22}\frac{I_2}{N_2}\right)-\varepsilon_2 \beta_2p_{22}S_2\frac{D_2}{N_2}\\\\
\dot E_2=\beta_1p_{21}S_2\left(p_{11}\frac{I_1}{N_1}+p_{21}\frac{I_2}{N_2}\right)+\beta_2p_{22}S_2\left(p_{12}\frac{I_1}{N_1}+p_{22}\frac{I_2}{N_2}\right) +\varepsilon_2 \beta_2p_{22}S_2\frac{D_2}{N_2} -\kappa E_2-\alpha E_2\\\\
\dot I_2=\kappa E_2-\gamma  I_2\\\\
\dot D_2=f_{\textrm{dead}}\gamma I_2-\nu D_2\\\\
\dot R_2=(1-f_{\textrm{dead}})\gamma I_2+\nu D_2+\alpha E_2
\end{array}\right.
\end{equation}
The difference, in the formulation of the infection term, from the one considered in \cite{bichara2015sis} is the \textit{effective} density of  infected. Here, the \textit{effective} density of infected  in Patch 1, for example, is  $$p_{11}\frac{I_1}{N_1}+p_{21}\frac{I_2}{N_2}$$ whereas in \cite{bichara2015sis}, it is $$\frac{p_{11}I_1+p_{21}I_1}{p_{11}N_1+p_{21}N_1}.$$
Focusing on  the changes on $E_1, \; I_1,\;D_1,\;E_2, \; I_2\;$ and $D_2$ and making use of the next generation approach we arrive at the basic reproductive number for the entire system, namely,
\begin{multline*}
$$\mathcal R_0=\dfrac{\kappa}{2(\kappa+\alpha)}\left(\dfrac{\beta_1p_{11}^2 +\beta_2p_{12}^2}{\gamma} +\dfrac{f_\textrm{death}\varepsilon_1\beta_1p_{11}}{\nu}+\dfrac{\beta_1p_{21}^2+\beta_2p_{22}^2}{\gamma}+\dfrac{f_\textrm{death}\varepsilon_2\beta_2p_{22}}{\nu}\right.\\\\
\left.+\sqrt{
\begin{aligned}
   & \left(\frac{\beta_1p_{11}^2 +\beta_2p_{12}^2}{\gamma} +\frac{f_\textrm{death}\varepsilon_1\beta_1p_{11}}{\nu}\right)^2+\left(\frac{\beta_1p_{21}^2+\beta_2p_{22}^2}{\gamma}+\frac{f_\textrm{death}\varepsilon_2\beta_2p_{22}}{\nu}\right)^2 \\
   & -2\left(\frac{\beta_1p_{11}^2 +\beta_2p_{12}^2}{\gamma} +\frac{f_\textrm{death}\varepsilon_1\beta_1p_{11}}{\nu}\right)\left(\frac{\beta_1p_{21}^2+\beta_2p_{22}^2}{\gamma}+\frac{f_\textrm{death}\varepsilon_2\beta_2p_{22}}{\nu}\right)\\
   & +4 \left(  \beta_1p_{11}p_{21} \frac{N_1}{\gamma N_2}+\beta_1p_{12}p_{22} \frac{N_1}{\gamma N_2} \right)\left( \beta_1p_{11}p_{21} \frac{N_2}{N_1}+\beta_1p_{12}p_{22} \frac{N_2}{N_1} \right)
\end{aligned}
}\right)
$$
\end{multline*}

We see, for example, that whenever the residents of Patch $j$ ($j=1,2$) live in communities where travel is not possible, that is, when $p_{12}=p_{21}=0$ or $p_{11}=p_{22}=1$, then the populations decouple and, consequently, we have that
$$\mathcal R_0=\max\{\mathcal R^1,\mathcal R^2\}$$
where $\displaystyle \mathcal R^i=\left(\frac{\beta_i}{\gamma}+\frac{1}{\nu}f_\textrm{death}\varepsilon_i\beta_i\right)\frac{\kappa}{\kappa+\alpha}$ for $i=1, 2$; that is, basic reproduction number of Patch $i$, $i=1,2$, if isolated.

\subsection{Final Epidemic Size in heterogeneous risk environments}
We keep track of the dead to make the mathematics simple. That is, to assuming that the population within each Patch is constant. And so, from the model, we get that

\begin{equation} \label{EboPatchyFin}
\left\{\begin{array}{ll}
\dot S_1=-\beta_1p_{11}S_1\left(p_{11}\frac{I_1}{N_1}+p_{21}\frac{I_2}{N_2}\right)-\beta_2p_{12}S_1\left(p_{12}\frac{I_1}{N_1}+p_{22}\frac{I_2}{N_2}\right)-\varepsilon_1 \beta_1p_{11}S_1\frac{D_1}{N_1}\\\\
\dot E_1=\beta_1p_{11}S_1\left(p_{11}\frac{I_1}{N_1}+p_{21}\frac{I_2}{N_2}\right)+\beta_2p_{12}S_1\left(p_{12}\frac{I_1}{N_1}+p_{22}\frac{I_2}{N_2}\right) +\varepsilon_1 \beta_1p_{11}S_1\frac{D_1}{N_1}-(\kappa+\alpha) E_1\\\\
\dot I_1=\kappa E_1-\gamma  I_1\\\\
\dot D_1=f_{\textrm{dead}}\gamma I_1-\nu D_1\\\\
\dot S_2=-\beta_1p_{21}S_2\left(p_{11}\frac{I_1}{N_1}+p_{21}\frac{I_2}{N_2}\right)-\beta_2p_{22}S_2\left(p_{12}\frac{I_1}{N_1}+p_{22}\frac{I_2}{N_2}\right)-\varepsilon_2 \beta_2p_{22}S_2\frac{D_2}{N_2}\\\\
\dot E_2=\beta_1p_{21}S_2\left(p_{11}\frac{I_1}{N_1}+p_{21}\frac{I_2}{N_2}\right)+\beta_2p_{22}S_2\left(p_{12}\frac{I_1}{N_1}+p_{22}\frac{I_2}{N_2}\right) +\varepsilon_2 \beta_2p_{22}S_2\frac{D_2}{N_2} -(\kappa+\alpha) E_2\\\\
\dot I_2=\kappa E_2-\gamma  I_2\\\\
\dot D_2=f_{\textrm{dead}}\gamma I_2-\nu D_2,
\end{array}\right.
\end{equation}

with initial conditions
$$S_1(0)=N_1,\quad E_1(0)=0,\quad I_1(0)=0,\quad D_1(0)=0,$$

$$S_2(0)=N_2,\quad E_2(0)=0,\quad I_2(0)=0,\quad D_2(0)=0,$$

We use the above model to find an ``approximate'' final size relationship.

\subsubsection*{Notation}

We make use of the notation $\hat g(t)$ for $\int_0^tg(s)ds$ and $g^\infty$ for $\lim_{t\to+\infty}g(t)$. We see that our analysis results guarantee that if $g(t)$ is a positive decreasing function then $g^\infty=0$.

Since $\dot S_1+\dot E_1=-(\kappa+\alpha)E_1\leq0$, then $E_1^\infty=0$ and since $\dot S_1+\dot E_1+I_1=-\alpha E_1-\gamma I_1\leq0$ then $I_1^\infty=0$. If we now consider that $\dot S_1+\dot E_1+I_1+D_1=-\alpha E_1-(1-f_{\textrm{dead}})\gamma I_1-\nu D_1\leq0$ then it follows that $D_1^\infty=0$. Similarly, it can be shown that  
$$E_2^\infty=I_2^\infty=D_2^\infty=0.$$
Focusing on the first two equations of System (\ref{EboPatchyFin}), we arrive at $$S_1^\infty-N_1=-(\kappa+\alpha) \hat E_1.$$
Consequently, since $\dot I_1=kE_1-\gamma I_1$, we have that $I_1^\infty=\kappa \hat E_1-\gamma \hat I_1$ and therefore $$\kappa \hat E_1=\gamma \hat I.$$
Using the equation for $\dot D_1$, we find that 
$$\nu\hat D_1=f_{\textrm{dead}}\gamma\hat I_1.$$
Similarly, we can deduce the analogous relationships for Patch 2, namely that,
$$S_2^\infty-N_2=-(\kappa+\alpha) \hat E_2,\quad \kappa \hat E_2=\gamma \hat I\quad\textrm{and}\quad  \nu\hat D_2=f_{\textrm{dead}}\gamma\hat I_2$$
From the equation for susceptible populations in Patch 1, we have that
$$\frac{\dot S_1}{S_1}=-\beta_1p_{11}\left(p_{11}\frac{I_1}{N_1}+p_{21}\frac{I_2}{N_2}\right)-\beta_2p_{12}\left(p_{12}\frac{I_1}{N_1}+p_{22}\frac{I_2}{N_2}\right)-\varepsilon_1 \beta_1p_{11}\frac{D_1}{N_1}$$
and, therefore that,
$$\log\frac{ S_1^0}{S_1^\infty}=\beta_1p_{11}\left(p_{11}\frac{\hat I_1}{N_1}+p_{21}\frac{\hat I_2}{N_2}\right)+\beta_2p_{12}\left(p_{12}\frac{\hat I_1}{N_1}+p_{22}\frac{\hat I_2}{N_2}\right)+\varepsilon_1 \beta_1p_{11}\frac{\hat D_1}{N_1}.$$
For the second patch, we have that
$$\log\frac{S_2^0}{S_2^\infty}=\beta_1p_{21}\left(p_{11}\frac{\hat I_1}{N_1}+p_{21}\frac{\hat I_2}{N_2}\right)+\beta_2p_{22}\left(p_{12}\frac{\hat I_1}{N_1}+p_{22}\frac{\hat I_2}{N_2}\right)+\varepsilon_2 \beta_2p_{22}\frac{\hat D_2}{N_2}.$$
Rewriting the expressions of $\hat I_i$ and $\hat D_i$ in terms of $S_i^\infty$,  $S_i^0$, $E_i^0$ and $I_i^0$, we arrive at the following two-patch ``approximate'' (since we are counting the dead), the final size relation. More precisely, with $N^0=N$, we have that
\begin{align*}
\log\dfrac{N_1}{S_1^\infty}=&\beta_1p_{11}\left(\dfrac{p_{11}\kappa}{\gamma(\kappa+\alpha)}\left(1-\frac{S_1^\infty}{N_1}\right)+\dfrac{p_{21}\kappa}{\gamma(\kappa+\alpha)}\left(1-\dfrac{S_2^\infty}{N_2}\right)\right)\\&+\beta_2p_{12}\left(\dfrac{p_{12}\kappa}{\gamma(\kappa+\alpha)}\left(1-\dfrac{S_1^\infty}{N_1}\right)+\dfrac{p_{22}\kappa}{\gamma(\kappa+\alpha)}\left(1-\dfrac{S_2^\infty}{N_2}\right)\right)\\
&+\varepsilon_1 \beta_1p_{11}\dfrac{f_{\textrm{dead}}}{\nu} \dfrac{\kappa}{\alpha+\kappa}\left(1-\dfrac{S_1^\infty}{ N_1}\right)
\end{align*}
\begin{align*}
\log\frac{N_2}{S_2^\infty}=&\beta_1p_{21}\left(\frac{p_{11}\kappa}{\gamma(\kappa+\alpha)}\left(1-\frac{S_1^\infty}{N_1}\right)+\frac{p_{21}\kappa}{\gamma(\kappa+\alpha)}\left(1-\frac{S_2^\infty}{N_2}\right)\right)
\\&+\beta_2p_{22}\left(\frac{p_{12}\kappa}{\gamma(\kappa+\alpha)}\left(1-\frac{S_1^\infty}{N_1}\right)+\frac{p_{22}\kappa}{\gamma(\kappa+\alpha)}\left(1-\frac{S_2^\infty}{N_2}\right)\right)\\
&+\varepsilon_2 \beta_2p_{22}\frac{f_{\textrm{dead}}}{\nu} \frac{\kappa}{\alpha+\kappa}\left(1-\frac{S_2^\infty}{ N_2}\right)
\end{align*}
Or in vectorial notation, we have that
{
\small{
\begin{equation}\label{FSEMatrix}
\begin{bmatrix}
   \log\dfrac{N_1}{S_1^\infty}\\\\
\log\dfrac{N_2}{S_2^\infty} 
  \end{bmatrix}
=
\begin{bmatrix}
K_{11} & K_{12} \\
\\
K_{21}& K_{22}
 \end{bmatrix}
 \begin{bmatrix}
1-\dfrac{S_1^\infty}{N_1}\\\\
1-\dfrac{S_2^\infty}{N_2} 
  \end{bmatrix}
\end{equation}
}}
where 
$$K_{11}=\left(\frac{\beta_1p_{11}^2 +\beta_2p_{12}^2}{\gamma} +\frac{f_\textrm{death}\varepsilon_1\beta_1p_{11}}{\nu}\right)\frac{\kappa}{\kappa+\alpha}.$$
Furthermore, we note that $K_{11}=A_1$ also appears in the next generation matrix, used to compute $\mathcal{R_0}$. Further, we also have that,
$$K_{12}=K_{21}=\left(  \beta_1 p_{11}p_{21}+\beta_2p_{12}p_{22}\right)\frac{\kappa}{\gamma(\kappa+\alpha)},$$
$$K_{22}=\left(\frac{\beta_1p_{21}^2+\beta_2p_{22}^2}{\gamma}+\varepsilon_2 \beta_2p_{22}\frac{f_{\textrm{dead}}}{\nu}\right) \frac{\kappa}{\alpha+\kappa}$$
Note that the vector in (\ref{FSEMatrix}) is given by
$$\begin{bmatrix}
1-\frac{S_1^\infty}{N_1}\\\\
1-\frac{S_2^\infty}{N_2} 
  \end{bmatrix}$$ 
representing the proportion of people in patches one and two able to transmit Ebola including transmission from handling dead bodies. $K_{12}^2=K_{12}K_{21}=A_2A_3$, $K_{22}=A_4$, we conclude that the matrix $K$ and the next generation matrix have the same eigenvalues, a result also found in  \cite{bichara2015sis}.

\section{Simulations}\label{sim2}
The basic model parameters used in the simulations are taken directly from the literature \cite{Legrand:2007ai,chowell2015modelling,TowersPattCCC2014,chowell2014transmission,AsympEbola2000} . We consider two patches and, for simplicity, it is assumed that they house the same number of individuals, namely, $N_1=N_2=1000000$. However, implicitly, it is assumed that the density is considerably higher in the high risk area. We assume that an outbreak starts in the high risk Patch 1 with $\beta_1=0.3056$. It propagates into Patch 2, low risk, defined by $\beta_2=0.1$. The difference between $\beta_1$ and $\beta_2$ or $\beta_1 - \beta_2$ provides a rough measure of the capacity to transmit, treat and control Ebola within connected two-patch systems. The initial conditions are set as  $S_{1}(0)=N-1,\; S_{2}(0)=N,\; E_{1,2}(0)=0,\;D_{1,2}(0)=0,\;R_{1,2}(0)=0,\;I_1=1,I_2=0$. The local basic reproductive numbers for each patch under isolation are $\mathcal{R}_0^1=2.41>1$ and $\mathcal{R}_0^2=1.08>1$.

We chose to report on three different mobility scenarios: one way movement, symmetric and asymmetric mobility. For the first case, only residents from Patch 1 travel, that is $p_{12}\geq 0$ and $p_{21}=0$. Given that Patch 1 is facing an epidemic, it is reasonable to assume that people in Patch 2 prefer to avoid traveling to Patch 1, and so, it is reasonable to assume that $p_{21}=0$. Mobility is allowed in both directions in a symmetric way, that is, residents of Patch 1 spend the same proportion of time in Patch 2 that individuals from Patch 2 spend in Patch 1; i.e. $p_{21}=p_{12}$. The third scenario assumes that mobility is asymmetric, and so, we make use, in this case, of the relation $p_{21}=1-p_{12}$.

\subsection{One way mobility}
\begin{figure}[H]
\centering
\includegraphics[scale=0.4]{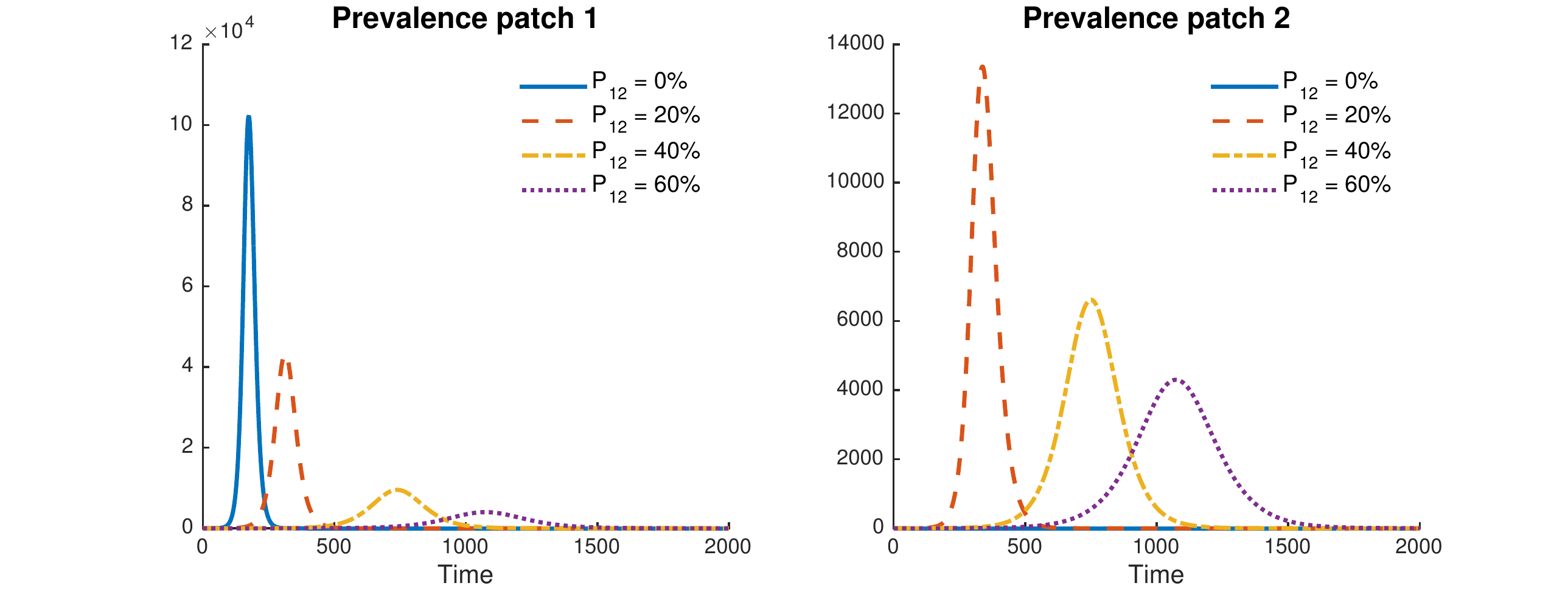}
\caption{Dynamics of prevalence in each Patch for values of mobility $p_{12}=0\%,\,20\%,\;40\%,\;60\%$ and $p_{21}=0$, with parameters: $\varepsilon_{1,2}=1, \beta_1=0.305, \beta_2=0.1, f_{death}=0.708, k=1/7, \alpha=0, \nu=1/2, \gamma=1/6.5$.}
\label{OWInfected}
\end{figure}

Simulations show that when only individuals from Patch 1 are allowed to travel, the prevalence and final size are lower that under a cordon sanitaire. Figure \ref{OWInfected}, shows the levels of Patch prevalence when $p_{12}=0\%,\,20\%,\,40\%$ and $60\%$. For low $p_{12}$'s, prevalence decreases in Patch 1 but remains high in both patches, which as expected, has a direct impact in the final size of the outbreak.

\begin{figure}[H]
\centering
\includegraphics[scale=0.3]{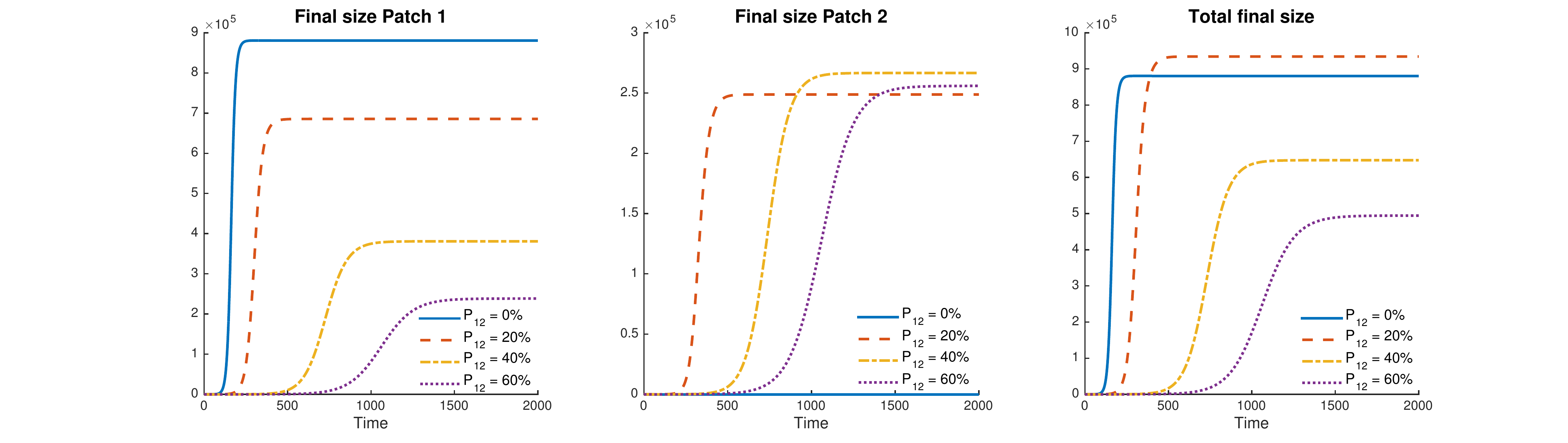}
\caption{Dynamics of prevalence in each Patch for values of mobility $p_{12}=0\%,\,20\%,\;40\%,\;60\%$ and $p_{21}=0$, with parameters: $\varepsilon_{1,2}=1, \beta_1=0.305, \beta_2=0.1, f_{death}=0.708, k=1/7, \alpha=0, \nu=1/2, \gamma=1/6.5$.}
\label{OWFinal}
\end{figure}

In Figure \ref{OWFinal}, simulations show that the total final size is only greater than the cordoned case when $p_{12}=20\%$, possibly the result of the assumption that $\gamma_1=\gamma_2$ and $\nu_1=\nu_2$. However, we see under the assumption of higher body disposal rates in Patch 2, that the total final size under $p_{12}=20\%$ may turn out to be smaller than in the cordoned case. That is, it is conceivable that a safer Patch 2, may emerge as a result of a better health care infrastructure and efficient protocols in the handling of dead bodies.

Finally, Figure \ref{OWFinal} shows that mobility can produce the opposite effect; that is, reduce the total final epidemic size, given that (for the parameters used) the residence times are greater than $p_{12}=25\%$ but smaller than $p_{12}=94\%$.
  \subsection{Symmetric mobility}

\begin{figure}[H]
\centering
\includegraphics[scale=0.4]{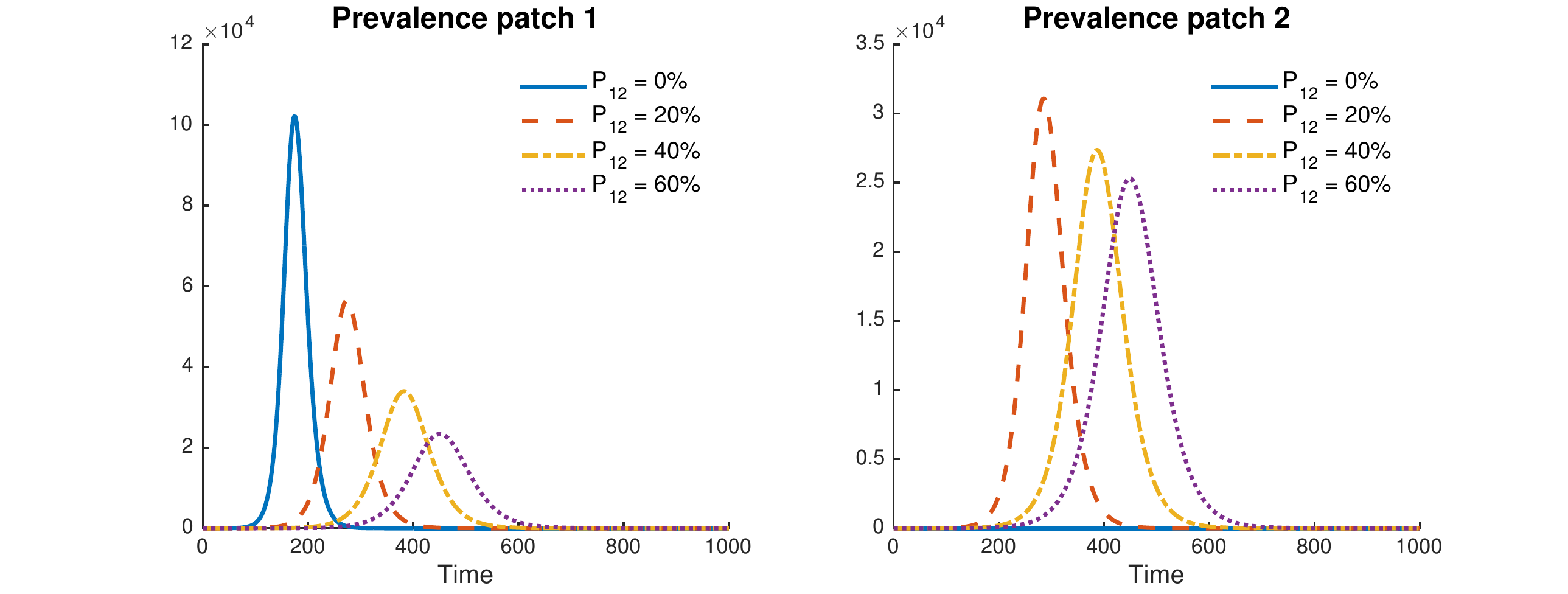}
\caption{Dynamics of prevalence in each Patch for values of mobility $p_{12}=0\%,\,20\%,\;40\%,\;60\%$ and $p_{21}=0$, with parameters: $\varepsilon_{1,2}=1, \beta_1=0.305, \beta_2=0.1, f_{death}=0.708, k=1/7, \alpha=0, \nu=1/2, \gamma=1/6.5$.}
\label{SInfected}
\end{figure}

Simulations under symmetric mobility show that prevalence and final size are severely affected when compared to the cordoned case. Figure \ref{SInfected} shows that 
the prevalence in Patch 1 exhibits the same behavior as in the one way scenario. However, in this case the prevalence in Patch 1 is decreasing at a slower rate due to the secondary infections produced by individuals traveling from Patch 2. On the other hand, prevalence in Patch 2 is much bigger than in the one way scenario, the result of secondary infections generated by individuals traveling from Patch 2 to Patch 1.

\begin{figure}[H]
\centering
\includegraphics[scale=0.3]{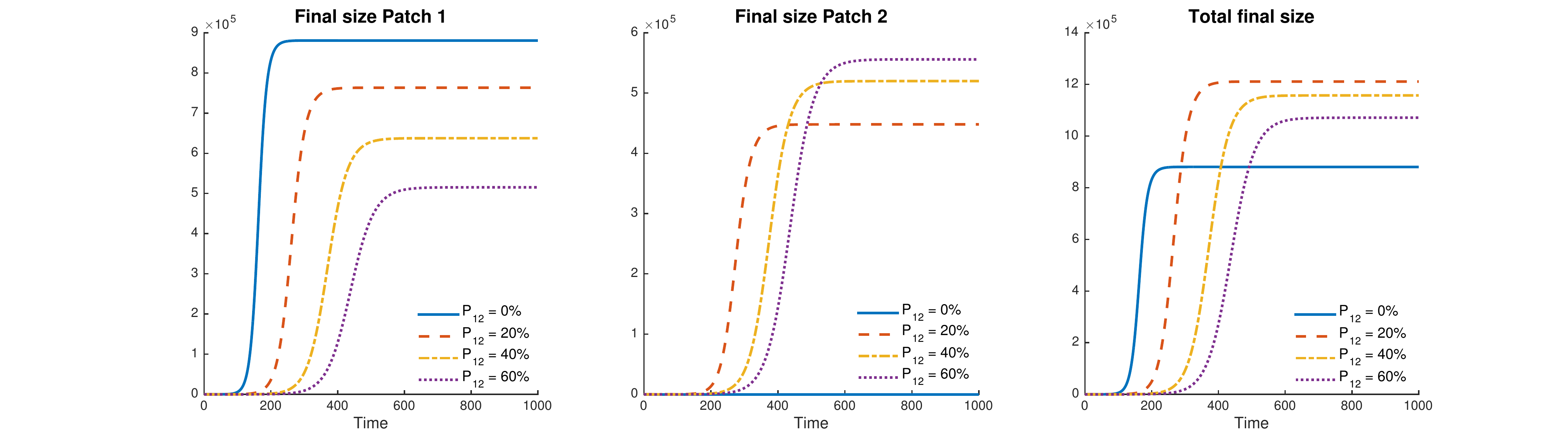}
\caption{Dynamics of prevalence in each Patch for values of mobility $p_{12}=0\%,\,20\%,\;40\%,\;60\%$ and $p_{21}=0$, with parameters: $\varepsilon_{1,2}=1, \beta_1=0.305, \beta_2=0.1, f_{death}=0.708, k=1/7, \alpha=0, \nu=1/2, \gamma=1/6.5$.}
\label{SFinal}
\end{figure}

We saw that final size in Patch 1 decreases when residency increases while an increment of the final size in Patch 2. That is, the total final size curve may turn out to be greater than in the cordoned case for almost all residence times. As seen in Figure \ref{SFinal}, allowing symmetric travel would negatively affect the total final size (almost always).

  \subsection{Final size analysis}
\begin{figure}[H]
\centering
\includegraphics[scale=0.4]{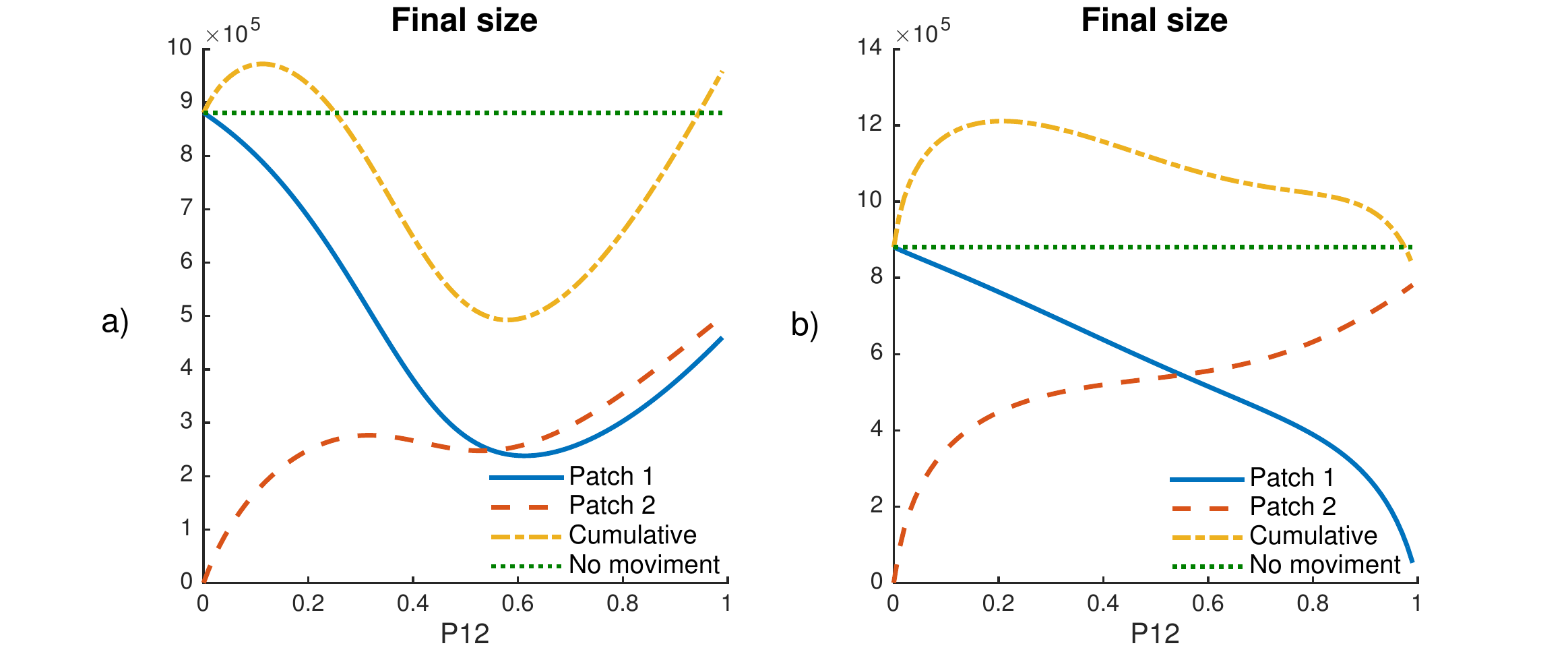}
\caption{Dynamics of maximum final size and maximum prevalence in Patch-one with parameters: $\varepsilon_{1,2}=1, \beta_1=0.305, \beta_2=0.1, f_{death}=0.708, k=1/7, \alpha=0, \nu=1/2, \gamma=1/6.5$.}
\label{maxonew}
\end{figure}

In order to clarify the effects of residence times and mobility on the total final size. We analyze its behavior under one way and symmetric mobility (Figure \ref{maxonew}). Figure \ref{maxonew}(a) shows,one way mobility, the existence of a proportional resident time interval when the total final size is reduced below that generated under the cordoned case. For residence times between $25\%$ and $94\%$. In particular, the best case scenario takes place when $p_{12}=58\%$, that is, when the final size reaches its all time minimum.

Figure \ref{maxonew}(b) shows that under symmetric mobility, the total final size increases for almost all resident times. Therefore traveling under these initial conditions has a deleterious effect to the overall population for almost all residence times.

  \subsection{Final size and basic reproductive number analysis}

\begin{figure}[H]
\centering
\includegraphics[scale=0.47]{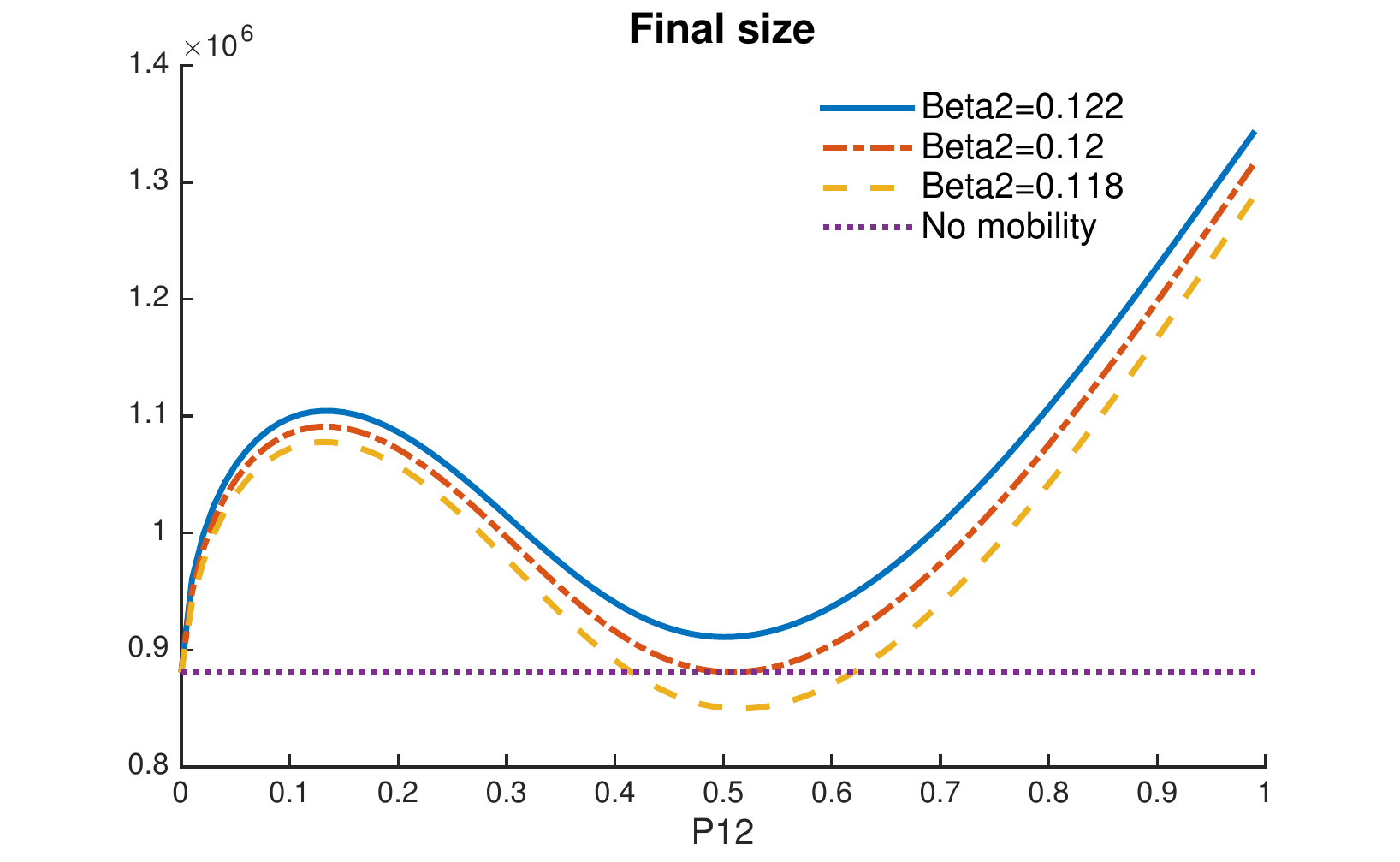}
\caption{Dynamics of maximum final size in the one way case with parameters: $\varepsilon_{1,2}=1, \beta_1=0.305, \beta_2=0.122,0.12,0.118, f_{death}=0.708, k=1/7, \alpha=0, \nu=1/2, \gamma=1/6.5$.}
\label{betas}
\end{figure}

It is important to notice that reductions in the total final size are related not only to residence times and mobility type but also to the prevailing infection rates. In Figure \ref{betas} simulations show the existence of an interval of residence times for which the total final size is less than the final size under the cordoned case under $\beta_2 < 0.12$. 

\begin{figure}[H]
\centering
\includegraphics[scale=0.4]{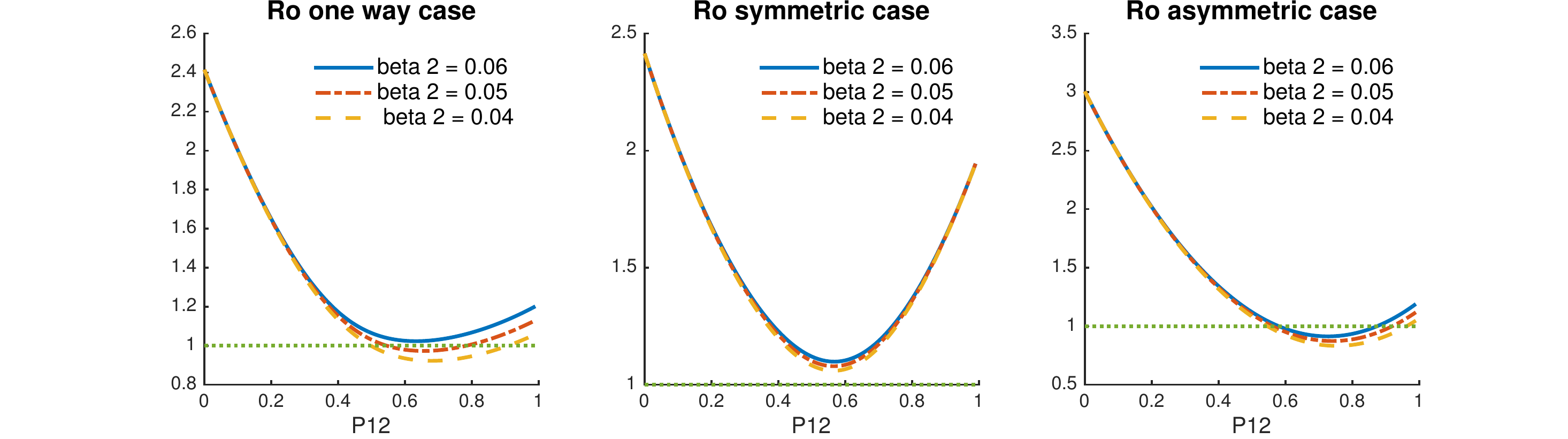}
\caption{Dynamics of $\mathcal R_0$ with parameters: $\varepsilon_{1,2}=1, \beta_1=0.305, \beta_2=0.06,0.05,0.04, f_{death}=0.708, k=1/7, \alpha=0, \nu=1/2, \gamma=1/6.5$.}
\label{Ro}
\end{figure}

Simulations (see Figure \ref{Ro}) show that mobility is always beneficial, that is, it reduces the global $\mathcal R_0$. However, mobility on its own is not enough to reduce $\mathcal R_0$ below the threshold (less than $1$). Bringing $\mathcal R_0<1$ would require reducing local risk, that is, getting a lower $\beta_2$.

\section{Conclusion}\label{Dis}
A West-Africa calibrated two-patch model of the transmission dynamics of EVD is used to show that the use of {\it cordons sanitaires} not always leads to the best possible global scenario and neither does allowing indiscriminate mobility. Mobility may reduced the total epidemic size as long as the low risk Patch 2 is ``safe enough'', otherwise mobility would produce a detrimental effect. Having an infection rate $\beta_2<0.12$ in Patch 2 guarantees (under our simulations) the existence of non-trivial residence times that reduce the total final size under one way mobility. The global basic reproductive number may be brought bellow one by mobility, whenever a the transmission rate in Patch 2 is low enough. Finally, the choice of non zero $\alpha$, that is, the recovery rate of asymptomatic that do not develop infection, bring the reproduction number $\mathcal R_0$ below one much faster for one way mobility than the case of $\alpha=0$  for a wide range of residence times. 
\subsection*{Acknoledgment:} \small{We want to thank Kamal Barley for providing us the figure \ref{lagmov}. These studies were made possible by grant \#1R01GM100471-01 from the National Institute of General Medical Sciences (NIGMS) at the National Institutes of Health. The contents of this manuscript are solely the responsibility of the authors and do not necessarily represent the official views of DHS or NIGMS. The funders had no role in study design, data collection and analysis, decision to publish, or preparation of the manuscript.}. 
\nocite{*}
\bibliographystyle{siam}


\begin{appendices}
\section{Appendix}
\subsection{Computation of $\mathcal R_0$}
Let us consider the infected compartments, i.e. $E, \; I$ and $D$. By following the next generation approach \cite{MR1057044,VddWat02}, we have that:

$$\mathcal F=\left(\begin{array}{c}
\beta S\frac{I}{N}+\varepsilon\beta S\frac{D}{N}\\
0\\
0
\end{array}\right)\quad\textrm{and}\quad \mathcal V=\left(\begin{array}{c}
-(\kappa+\alpha) E\\
\kappa E-\gamma I\\
f_{\textrm{dead}}\gamma I-\nu D
\end{array}\right)$$
thus, we have:
$$\mathcal {DF}=\left(\begin{array}{ccc}
0 & \beta \frac{S}{N} &\varepsilon\beta \frac{S}{N}\\
0 & 0 & 0\\
0 & 0 & 0
\end{array}\right)\quad\textrm{and}\quad \mathcal DV=\left(\begin{array}{ccc}
-(\kappa+\alpha)  & 0 & 0\\
\kappa &-\gamma & 0\\
0 &  f_{\textrm{dead}}\gamma &-\nu 
\end{array}\right).$$
At the DFE, $S=N$, hence
$$F=\left(\begin{array}{ccc}
0 & \beta  & \varepsilon \beta \\
0 & 0 & 0\\
0 & 0 & 0
\end{array}\right)\quad\textrm{and}\quad V=\left(\begin{array}{ccc}
-(\kappa+\alpha)  & 0 & 0\\
\kappa &-\gamma & 0\\
0 &  f_{\textrm{dead}}\gamma &-\nu 
\end{array}\right),$$
and the basic reproduction number is the spectral radius of the next generation matrix:
$$
-FV^{-1}=\left(\begin{array}{ccc}
\frac{\kappa\beta}{(\kappa+\alpha)\gamma} +\frac{\varepsilon\kappa f_{\textrm{dead}}\beta}{(\kappa+\alpha)\nu}  & \frac{\beta}{\gamma} +\frac{\varepsilon f_{\textrm{dead}}\beta}{\nu} & \frac{\varepsilon\beta}{\nu} \\
0 &0& 0\\
0 &  0 &0 
\end{array}\right)
.$$

Thus the basic reproduction number is
$$\mathcal R_0=\left(\frac{\beta}{\gamma}+\frac{\varepsilon f_{\textrm{dead}}\beta}{\nu}\right)\frac{\kappa}{\kappa+\alpha},$$

\subsection{Final Epidemic Size and Exponential growth rates}
The total population of system (\ref{EbolaAsym}) is constant, we can consider only the system
\begin{equation}
\left\{\begin{array}{ll}
\dot S=-\beta S\frac{I}{N}-\varepsilon\beta S\frac{D}{N}\\\\
\dot E=\beta S\frac{I}{N}+\varepsilon_D\beta S\frac{D}{N}-(\kappa+\alpha) E\\\\
\dot I= \kappa E-\gamma I\\\\
\dot D=f_{\textrm{dead}}\gamma I-\nu D
\end{array}\right.
\end{equation}
We suppose $S(0)=N, E(0)=I(0)=D(0)=0$. By summing the first two equations of (\ref{Ebola2}), we have: $\dot S+\dot E=-(\kappa+\alpha)E\leq0$. This implies that $E^\infty=0$. Similarly by adding the first three and first four equations, we will have $I^\infty=0$ and $D^\infty=0$.\\

By integrating the first 2 equations, we have $\displaystyle S^\infty-N=-(\kappa+\alpha)\hat E$. Hence $\displaystyle\hat E=\frac{N-S^\infty}{\kappa+\alpha}$

Similarly, we have $\displaystyle \hat I=\frac{\kappa}{\gamma(\kappa+\alpha)}(N-S^\infty)$ and $\displaystyle \hat D=\frac{f_{\textrm{dead}}}{\nu}\frac{\kappa}{\kappa+\alpha}(N-S^\infty$\\
By using the first equation, we have:

$$\log\frac{N}{S^\infty}=\frac{\beta}{\gamma}\frac{\kappa}{\kappa+\alpha}\frac{N-S^\infty}{N}+\varepsilon \beta\frac{f_{\textrm{dead}}}{\nu}\frac{\kappa}{\kappa+\alpha}\frac{N-S^\infty}{N}$$
Hence, we have the final epidemic relation:
$$\log\frac{N}{S^\infty}=\mathcal R_0\left(1-\frac{S^\infty}{N}\right)$$

\subsection{Computation of $R_0$ in heterogeneous risk environments}
In heterogeneous risk environments let us consider the infected compartments, i.e. $E_1, \; I_1,\;D_1,\;E_2, \; I_2\;$ and $D_2$. By following the next generation approach, we have:

$$\mathcal F=\left(\begin{array}{c}
\beta_1p_{11}S_1\left(p_{11}\frac{I_1}{N_1}+p_{21}\frac{I_2}{N_2}\right)+\beta_2p_{12}S_1\left(p_{12}\frac{I_1}{N_1}+p_{22}\frac{I_2}{N_2}\right) +\varepsilon_1 \beta_1p_{11}S_1\frac{D_1}{N_1}\\
0\\
0\\
\beta_1p_{21}S_2\left(p_{11}\frac{I_1}{N_1}+p_{21}\frac{I_2}{N_2}\right)+\beta_2p_{22}S_2\left(p_{12}\frac{I_1}{N_1}+p_{22}\frac{I_2}{N_2}\right) +\varepsilon_2 \beta_2p_{22}S_2\frac{D_2}{N_2}\\
0\\
0
\end{array}\right)$$
And 
$$ \mathcal V=\left(\begin{array}{c}
-(\kappa+\alpha) E_1\\
\kappa E_1-\gamma  I_1\\
f_{\textrm{dead}}\gamma I_1-\nu D_1\\
-(\kappa+\alpha) E_2\\
\kappa E_2-\gamma  I_2\\
f_{\textrm{dead}}\gamma I_2-\nu D_2\\
\end{array}\right)$$
Hence, we have:
$$\mathcal {DF}=\left(\begin{array}{cccccc}
0 & \beta_1p_{11}^2 \frac{S_1}{N_1}+\beta_2p_{12}^2 \frac{S_1}{N_1}  & \beta_1p_{11}\varepsilon_1 \frac{S_1}{N_1}&0&\beta_1p_{11}p_{21} \frac{S_1}{N_2}+\beta_11p_{12}p_{22} \frac{S_1}{N_2}&0\\
0 & 0 & 0 &0&0&0\\
0 & 0 & 0 &0&0&0\\
0 & \beta_1p_{11}p_{21} \frac{S_2}{N_1}+\beta1p_{12}p_{22} \frac{S_2}{N_1} & 0 &0&\beta_1p_{21}^2 \frac{S_2}{N_2}+\beta_2p_{22}^2 \frac{S_2}{N_2}&\beta_2p_{22}\varepsilon_2 \frac{S_2}{N_2}\\
0 & 0 & 0 &0&0&0\\
0 & 0 & 0 &0&0&0
\end{array}\right)$$ 
and
$$ \mathcal {DV}=\left(\begin{array}{cccccc}
-(\kappa+\alpha)  & 0 & 0 &0&0&0\\
\kappa  & -\gamma & 0 &0&0&0\\
0  & f_\textrm{death}\gamma & -\nu &0&0&0\\
0 & 0 & 0 &-(\kappa+\alpha)&0&0\\
0 & 0 & 0 &\kappa&-\gamma&0\\
0  & 0 & 0 &0&f_\textrm{death}\gamma&-\nu
\end{array}\right)$$

At the DFE, $S_1^*=N_1$ and $S_2^*=N_2$, hence
$$F=\left(\begin{array}{cccccc}
0 & \beta_1p_{11}^2 +\beta_2p_{12}^2  & \beta_1p_{11}\varepsilon_1 &0&\beta_1p_{11}p_{21} \frac{N_1}{N_2}+\beta_11p_{12}p_{22} \frac{N_1}{N_2}&0\\
0 & 0 & 0 &0&0&0\\
0 & 0 & 0 &0&0&0\\
0 & \beta_1p_{11}p_{21} \frac{N_2}{N_1}+\beta1p_{12}p_{22} \frac{N_2}{N_1} & 0 &0&\beta_1p_{21}^2+\beta_2p_{22}^2&\beta_2p_{22}\varepsilon_2\\
0 & 0 & 0 &0&0&0\\
0 & 0 & 0 &0&0&0
\end{array}\right)$$
and
$$ V=\left(\begin{array}{cccccc}
-(\kappa+\alpha)  & 0 & 0 &0&0&0\\
\kappa  & -\gamma & 0 &0&0&0\\
0  & f_\textrm{death}\gamma & -\nu &0&0&0\\
0 & 0 & 0 &-(\kappa+\alpha)&0&0\\
0 & 0 & 0 &\kappa&-\gamma&0\\
0  & 0 & 0 &0&f_\textrm{death}\gamma&-\nu
\end{array}\right)$$
The basic reproduction number is the spectral radius of the next generation matrix:
$$
-FV^{-1}=\left(\begin{array}{cccccc}
A_1& A_2 &\frac{ \beta_1p_{11}\varepsilon_1}{\nu} &A_3&A_4&0\\
0 & 0 & 0 &0&0&0\\
0 & 0 & 0 &0&0&0\\
A_5 &A_6 & 0 &A_7&A_8&\frac{\beta_2p_{22}\varepsilon_2}{\nu}\\
0 & 0 & 0 &0&0&0\\
0 & 0 & 0 &0&0&0
\end{array}\right)$$
where 
\begin{align*}
A_1&=\left(\frac{\beta_1p_{11}^2 +\beta_2p_{12}^2}{\gamma} +\frac{f_\textrm{death}\varepsilon_1p_{11}\beta_1}{\nu}\right)\frac{\kappa}{\kappa+\alpha},\\\\ 
A_2&=\frac{\beta_1p_{11}^2 +\beta_2p_{12}^2}{\gamma} +\frac{f_\textrm{death}\varepsilon_1\beta_1p_{11}}{\nu},\\\\
A_3&=\left(\beta_1p_{11}p_{21}+\beta_2p_{12}p_{22} \right)\frac{N_1}{N_2}\frac{\kappa}{\gamma(\kappa+\alpha)},\\\\
A_4&=(\beta_1p_{11}p_{21}+\beta_2p_{12}p_{22}) \frac{N_1}{\gamma N_2},\\\\
A_5&=\left(\beta_1p_{11}p_{21}+\beta_2p_{12}p_{22} \right)\frac{N_2}{N_1}\frac{\kappa}{\gamma(\kappa+\alpha)}=\left(\frac{N_2}{N_1}\right)^2A_3,\\\\
A_6&=\frac{1}{\gamma}\left(\beta_1p_{11}p_{21}+\beta_2p_{12}p_{22} \right)\frac{N_2}{N_1},\\\\
A_7&=\left(\frac{\beta_1p_{21}^2+\beta_2p_{22}^2}{\gamma}+\frac{f_\textrm{death}\varepsilon_2\beta_2p_{22}}{\nu}\right)\frac{\kappa}{\kappa+\alpha},\\\\
A_8&=\frac{\beta_1p_{21}^2+\beta_2p_{22}^2}{\gamma}+\frac{f_\textrm{death}\varepsilon_2\beta_2p_{22}}{\nu}.
\end{align*}

We can easily see that $-FV^{-1}$ has the same nonzero eigenvalues as the matrix

$$
\left(\begin{array}{cc}
A_1& A_3\\
A_5  & A_7
\end{array}\right)=\left(\begin{array}{cc}
\tilde A_1& \tilde A_2\\
\tilde A_3  & \tilde A_4
\end{array}\right)$$

\begin{eqnarray}
\mathcal R_0&=&\frac{1}{2}\left(\tilde A_1+\tilde A_4+\sqrt{(\tilde A_1+\tilde A_4)^2-4(\tilde A_1\tilde A_4-\tilde A_2\tilde A_3)} \right)\frac{\kappa}{\kappa+\alpha}\nonumber\\
&= &\frac{1}{2}\left( \tilde A_1+\tilde A_4+\sqrt{\tilde A_1^2+\tilde A_4^2+2\tilde A_1\tilde A_4-4(\tilde A_1\tilde A_4-\tilde A_2\tilde A_3)} \right)\nonumber\\
&= &\frac{1}{2}\left( \tilde A_1+\tilde A_4+\sqrt{ \tilde A_1^2+\tilde A_4^2-2\tilde A_1\tilde A_4+4\tilde A_2\tilde A_3} \right)\nonumber
\end{eqnarray}
More precisely, we have:

\begin{multline*}
$$\mathcal R_0=\dfrac{\kappa}{2(\kappa+\alpha)}\left(\dfrac{\beta_1p_{11}^2 +\beta_2p_{12}^2}{\gamma} +\dfrac{f_\textrm{death}\varepsilon_1\beta_1p_{11}}{\nu}+\dfrac{\beta_1p_{21}^2+\beta_2p_{22}^2}{\gamma}+\dfrac{f_\textrm{death}\varepsilon_2\beta_2p_{22}}{\nu}\right.\\\\
\left.+\sqrt{
\begin{aligned}
   & \left(\frac{\beta_1p_{11}^2 +\beta_2p_{12}^2}{\gamma} +\frac{f_\textrm{death}\varepsilon_1\beta_1p_{11}}{\nu}\right)^2+\left(\frac{\beta_1p_{21}^2+\beta_2p_{22}^2}{\gamma}+\frac{f_\textrm{death}\varepsilon_2\beta_2p_{22}}{\nu}\right)^2 \\
   & -2\left(\frac{\beta_1p_{11}^2 +\beta_2p_{12}^2}{\gamma} +\frac{f_\textrm{death}\varepsilon_1\beta_1p_{11}}{\nu}\right)\left(\frac{\beta_1p_{21}^2+\beta_2p_{22}^2}{\gamma}+\frac{f_\textrm{death}\varepsilon_2\beta_2p_{22}}{\nu}\right)\\
   & +4 \left(  \beta_1p_{11}p_{21} \frac{N_1}{\gamma N_2}+\beta_1p_{12}p_{22} \frac{N_1}{\gamma N_2} \right)\left( \beta_1p_{11}p_{21} \frac{N_2}{N_1}+\beta_1p_{12}p_{22} \frac{N_2}{N_1} \right)
\end{aligned}
}\right)
$$
\end{multline*}
\end{appendices}
\end{document}